
\input phyzzx
\hoffset=0.375in
\overfullrule=0pt

\twelvepoint
\font\bigfont=cmr17
\centerline{\bigfont The Mass Spectrum Of Machos}
\centerline{\bigfont From Parallax Measurements}
\bigskip
\centerline{\bf Cheongho Han}
\smallskip
\centerline{\bf Andrew Gould}
\smallskip
\centerline{Dept of Astronomy, Ohio State University, Columbus, OH 43210}
\bigskip
\centerline{e-mail cheongho@payne.mps.ohio-state.edu}
\centerline{e-mail gould@payne.mps.ohio-state.edu}
\bigskip
\centerline{\bf Abstract}

\doublespace

     We demonstrate that by making satellite-based parallax measurements of
Macho
events, it is possible to distinguish disk from bulge Machos and to determine
individual masses of disk Machos to an accuracy of $0.2$ in the log.
This is at least a 3-fold improvement over what can be done without parallaxes,
i.e.,
just from the observed time scale of the events.
In addition, we show that the physical distribution
of disk Machos can be found from `reduced' velocities determined by parallax
measurements.

\endpage

\chapter{Introduction}

    Major efforts to detect dark matter by observing gravitationally
lensed stars located in the Galactic bulge are being carried out
by the OGLE (Udalski et al.\ 1994) and MACHO (Alcock et al.\ 1994).
Several dozen candidate events have already been reported.
The light curve of a lensing event is given by
$$
A\left[u(t)\right] = { u^{2} + 2 \over u(u^{2}+4)^{1/2} },
\ \ \ \ u(t) = \sqrt { \beta^{2} + \omega^{2} ( t - t_{0} )^{2} },
\eqno(1.1)
$$
where $\omega^{-1}$ is the characteristic time, $\beta$ is the dimensionless
impact parameter, and $t_{0}$ is the midpoint of the event.
Of the three parameters in equation (1.1), only $\omega^{-1}$ provides
information about the lenses responsible for the events.
The characteristic time is related to the physical parameters of the lens by
$$
\omega^{-1} = {   \left(4GMD_{ol}D_{ls}/D_{os}\right)^{1/2} \over
vc    },
\eqno(1.2)
$$
where $M$ and $v$ are the mass and transverse speed of the lens, and
$D_{ol}, D_{os}$, and $ D_{ls}$
are the distances between the observer, lens, and  source.
However, these physical parameters cannot be measured separately, and
hence the physical information about the lenses is poorly determined.

    There have been two general approaches advanced for finding additional
information about individual events.
The first one is to measure space-based parallaxes, which yield
``reduced'' transverse velocities $\tilde{{\bf v}}$ and corresponding
``reduced'' Einstein ring radius $\tilde{R_{e}}$ (Gould 1994b, 1994d),
where
$$
\tilde{{\bf v}} = { D_{os} \over D_{ls} } {\bf v},
\eqno(1.3)
$$
and
$$
\tilde{R}_{e} = { D_{os} \over D_{ls} }R_{e},\ \ \
R_{e} = \left( { 4GM\over c^{2} }
{D_{ol}D_{ls} \over D_{os} } \right)^{1/2}.
\eqno(1.4)
$$
When a single satellite is used for the measurement, there would generally
exist a four-fold degeneracy of $\tilde {{\bf v} }$; two-fold by magnitude
and the other two-fold by direction.
This degeneracy can be completely broken by the measurement of parallaxes from
a second satellite (Gould 1994b).
However, for most Macho events a single-satellite parallax is sufficient
to measure $\tilde{v}$ and $\tilde{R}_{e}$, leaving only a
two-fold degeneracy in direction (Gould 1994d).
The other approach is to measure the lens proper motion $\Omega = v/D_{ol}$
by photometric (Gould 1994a; Nemiroff \& Wickramasinghe 1994) or
spectroscopic (Maoz $\&$ Gould 1994) methods.

     In this paper we demonstrate that the events generated by bulge
and disk Machos could be distinguished and the individual masses of
disk Machos could be measured with a significantly improved accuracy
from the reduced speed measurements of events seen toward the galactic bulge.
Since the required parameter is the reduced speed not the velocity,
this method eliminates the need to launch a costly second satellite
for the complete measurements of $\tilde{{\bf v}}$.
We also discuss a method to extract information about
the physical distribution of disk Machos from the measured reduced speeds.
By contrast, we find that it is only rarely possible to measure proper motions
of Machos seen toward the bulge.
Moreover, even when they can be measured, proper motions in this direction
provide less information about the Machos than parallaxes.

     The reason that parallax measurements are sensitive to the mass of
disk Machos while proper motions are not, can be understood as follows.
The flat rotation of the disk implies that the mean transverse velocity
as a function of distance, $\bar {{\bf v}}(D_{ol})$, is given by
$$
\bar{{\bf v}}(D_{ol}) = {D_{ol} \over R_{0}} (v_{\odot},0),
\eqno(1.5)
$$
where $v_{\odot} = 220\ {\rm km\ s^{-1}}$ is the rotation speed of the Sun and
$R_{0} = 8\ {\rm kpc}$ is the solar galactocentric distance.
Hence, the mean reduced speed is a definite function of $D_{ol}$.
Measurement of $\tilde{v}$ therefore strongly constrains $D_{ol}$.
To the extent that $\tilde{v}$ and $D_{ol}$ are known, the mass is
also known.
However, from equation (1.5), the mean angular velocity
$\overline{ {\bf \Omega} }(D_{ol}) = (v_{\odot}/R_{0}, 0)$ is independent of
distance.
Hence, a proper motion measurement provides almost no information about the
distance.

\chapter{Probability Of Lensing Events}
\section{Bulge Lensing Event}

    The optical depth to lensng is defined as the angular area covered by the
Einstein ring $A=\pi (R_{e} / D_{ol})^{2}$ as a fraction of the sky.
If the bulge stars are distributed over distances between $d_{1}$ and $d_{2}$
from the Earth with a number density distribution $n(D_{os})$,
the optical depth $\tau$ of a bulge star lensed by foreground
bulge stars is
$$
\tau_{bulge} = {4\pi G \over c^{2}}
{\int_{d_{1}}^{d_{2}}dD_{os}n(D_{os})\int_{d_{1}}^{D_{ol}}
dD_{ol}  \rho (D_{ol}) D }
\left[ \int_{d_{1}}^{d_{2}} dD_{os}n(D_{os}) \right]^{-1},
\eqno(2.1.1)
$$
where $D \equiv D_{ol}D_{ls} / D_{os}$, and
$\rho (D_{ol})$ is the mass density of Machos along the line of sight.
As the distance from the observer increases, the volume element and total
number of
stars in the element increase.
However, the actual number of stars one can detect decreases with increasing
distance due to detection limits.
In our computation, we assume that these two factors cancel out.
This is equivalent to the $\beta = -1$ model of Kiraga \& Paczy\'nski
(1994, hereafter KP).

     The frequency of lensing events lensed by the bulge stars,
$\Gamma_{bulge}$, is
$$
\Gamma_{bulge}=4\sqrt{{ G\over c^{2}M }}
\int_{d_{1}}^{d_{2}} dD_{os} n(D _{os})
\int_{d_{1}}^{D_{ol}} dD_{ol} \rho (D_{ol}) D^{1/2}
$$
$$
\times
\int dv_{y} \int dv_{z}vf(v_{y},v_{z})
\ \left[ \int_{d_{1}}^{d_{2}} dD_{os} n(D_{os}) \right]^{-1},
\eqno(2.1.2)
$$
where $v_{y}$ and $v_{z}$ are the components of ${\bf v}$
and $f(v_{y},v_{z})$ is their distribution.
Here the coordinates $(x,y,z)$ have their center at the center of the Galaxy
and $x$ and $z$ axes point to the Earth and to the north galactic pole,
respectively.

\section{Disk Lensing Events}

     The source bulge stars are located within a narrow region compared
to the typical values of $D_{ol}$ and $D_{ls}$ of disk lenses.
Therefore, it is possible to approximate the source stars as being located at
a fixed distance.
We assume $D_{os}= 8\ {\rm kpc}$.
Then the optical depth is
$$
\tau_{disk} = {4\pi G \over c^{2}} \int_{0}^{d_{2}} dD_{ol}
\rho (D_{ol})  D.
\eqno(2.2.1)
$$

     Similarily, the event rate of a bulge star being lensed by the disk stars,
$\Gamma_{disk}$, is
$$
\Gamma_{disk}=4\sqrt{{ G\over c^{2}M }}
\int_{0}^{D_{os}} dD_{ol} \rho (D_{ol}) D^{1/2}
\int  dv_{y} \int dv_{z} vf(v_{y},v_{z}).
\eqno(2.2.2)
$$

\chapter{Models}

     To compute $\tau$ and $\Gamma$, we adopt various
models of velocity and density distributions.
The results are used to investigate the sensitivity of $\tau$ and $\Gamma$
to the choice of model.
We always assume that the number density of the bulge sources is proportional
to mass density of the bulge lenses, $n \propto \rho$.

     The transverse velocity ${\bf v}$ is defined as
$$
{\bf v} =  {\bf v}_{l} -  {\bf v}_{los} =  {\bf v}_{l} -
\left[  {\bf v}_{s}{ D_{ol}\over D_{os} }
+ {\bf v}_{o} { D_{ls}\over D_{os} } \right],
\eqno(3.1)
$$
where ${\bf v}_{l},{\bf v}_{s}$, and ${\bf v}_{o}$ are the transverse
velocities of the lens, the source, and the observer, respectively.
We assume that the velocity distribution is gaussian
$f(v_{y},v_{z}) = f(v_{y})f(v_{z})$, where
$$
f(v_{y}) = { 1\over \sqrt{2\pi \sigma_{y}^{2}} }
\exp \left[ -{ (v_{y}-\bar v_{y})^{2} \over  2\sigma_{y}^{2} } \right],
\eqno(3.2)
$$
and similarily for $f(v_{z})$.
We adopt $\bar v_{z,disk} = \bar v_{z,bulge} = 0$ and
$\sigma_{z,disk} = 20\ {\rm km\ s^{-1} },
\sigma_{z,bulge} = 100\ {\rm km\ s^{-1} }$ for the $z$ component of velocity.
For the $y$ direction, the values adopted are
$\bar v_{y,disk} = 220\ {\rm km\ s^{-1} }$ and
$\sigma_{y,disk} = 30\ {\rm km\ s^{-1} }$ for the disk lenses, whereas
$\bar v_{y,bulge} = 0$ and
$\sigma_{y,bulge} = 100\ {\rm km\ s^{-1} }$ for the bulge lenses.
Another disk model where the velocity dispersion increases steadily
toward the galactic center (Lewis \& Freeman 1989) is separately considered.
In this model (linear disk velocity model), the velocity dispersion in the
solar neighborhood is $(\sigma_{y,disk},
\sigma_{z,disk)} = (30,20)\ {\rm km\ s^{-1}}$, and
it increases linearly toward the galactic center until it becomes
$(\sigma_{y,disk}, \sigma_{z,disk)} = (75,50)\ {\rm km\ s^{-1}}$
at the galactic center.

     The influence of bulge rotation is tested separately by assuming that
$$
v_{rot} = v_{max} \left(x \over 1\ {\rm kpc} \right)\ \ \
(R<1 \ {\rm kpc,\  rigid\ body\ rotation}),
$$
$$
v_{rot} = v_{max} \left(x\over R\right)\ \ \
(R \geq 1\ {\rm kpc}, \ {\rm flat\ rotation}),
\eqno(3.3)
$$
where $v_{max}= 100\ {\rm km\ s^{-1}}$, and $R = \sqrt{x^{2} + y^{2} }$.

     We also consider the possibility of a bar-like structure for the
bulge and its influence on lensing events.
In the model the bar has axis ratio (1 : 0.33 : 0.23) and it
is rotated in the galactic plane with its near side in the first galactic
quadrant making an angle $\theta = 20^{\circ}$ between its major axis and
the line-of-sight to the galactic center (Dwek et al.\ 1994).
The velocity distribution for the barred bulge is deduced from the
tensor virial theorem (Binney $\&$ Tremaine 1987) with result
$(\sigma_{x'},\sigma_{y'},\sigma_{z'}) = (113.6,77.4,66.3)\ {\rm km\ s^{-1}}$.
Here, the coordinates $(x',y',z')$ have their center at the galactic center,
$x'$ represents the longest axis, and the shortest axis ($z'$) is
directed toward the north galactic pole.
Because the mass-to-light ratio of the bulge is not known {\it a priori},
the virial theorem gives only the axis ratios of the velocity
dispersion tensor.
To obtain our result, we normalize to the observed mean line-of-sight
velocity dispersion of
$\sim 110\ {\rm km\ s^{-1}}$.
This calculation yields a total bulge mass
$M_{bulge} \sim 1.2 \times 10^{10}\ {\rm M_{\odot}} $,
very close to Dwek et al.'s estimate of
$M_{bulge} \sim 1.3 \times 10^{10}\ {\rm M_{\odot}}$ based on
a Salpeter mass function cut off at $M = 0.1\ {\rm M_{\odot}}$.
The projected velocity dispersions are computed by
$\sigma^{2}_{x} = \sigma^{2}_{x'} \cos^{2} \theta + \sigma^{2}_{y'} \sin^{2}
\theta$,
$\sigma^{2}_{y} = \sigma^{2}_{x'} \sin^{2} \theta + \sigma^{2}_{y'} \cos^{2}
\theta$,
and $\sigma_{z} = \sigma_{z'}$.
The resultant velocity dispersions in the $(x,y,z)$ coordinates are
$(\sigma_{x},\sigma_{y},\sigma_{z}) = (110,82.5,66.3)\ {\rm km\ s^{-1}}$.
For the detailed computation of the optical depth and event rate
based on the barred bulge model, see Han (1994).
The transverse velocity distributions for various models
are listed in Table 1.

    There is disagreement over the three-dimensional distribution
of stellar populations of the Milky Way.
Therefore, it is useful to compare the effects on the optical depth,
event rate, and reduced velocities using various models.
We adopt three bulge models, isothermal, barred, and Kent (1992), and
three disk models, Bahcall (1986), Kent (1991), and
KP (1994).
The density distributions of the models are listed in Table 2.
Since the mass-to-light ratio of the disk stellar population are
not well determined, we normalize all disk models to the
mass density of solar neighborhood $\rho_{0} \sim 0.06\ {\rm M_{\odot}
pc^{-3}}$.
Note that this value is somewhat higher than the observed stellar mass density
near the Sun because the models are aimed at reproducing the star
distribution $\gsim 1$ scale height from the plane, not the local density.
The density profile through Baade's Window at $l = 1^{\circ},
b = -3^{\circ}\hskip-2pt .9 $
is shown with respect to the distance from the Earth in Figure 1.

     Because various researchers have adopted different models, they find
different optical depths and probabilities of lensing events from one another.
To see the effects on the optical depths of different mass models, we
compute the optical depths for various models and the results are listed
and compared with values estimated by others in Table 3.
The computed optical depths for various models are similar to each
other and are not strongly model-dependent except for the barred bulge model:
a barred bulge doubles the predicted bulge optical depth.

\chapter{Lens Locations and Masses}
\section{``Reduced'' Velocity}

     It would be very hard to untangle the contributions of
bulge and disk Machos from the analysis of $\omega^{-1}$ alone.
Griest et al.\ (1991) and KP (1994)
point out that the mean duration of events $\omega^{-1}$ would be longer
for lenses in the disk than for ones in the bulge.
However, this occurs mainly because the event-rate distribution for disk
events has a large tail toward the longer events.
By contrast, the most common characteristic time of events is almost the same
regardless of whether the lenses are bulge or disk Machos making it difficult
to distinguish between them (Evans 1994).
The value of $\omega^{-1}$ for the most common events is
$\sim 7\ ({\rm M}/0.1\ {\rm M_{\odot}})^{1/2}$
days, for both bulge and disk Machos.

     However, it would be possible to distinguish the bulge and disk Machos
if their reduced velocities could be measured.
The rates of lensing events by bulge and disk stars are computed
from the equations in \S\ 2 and are presented with respect to
$\tilde{v}_{y}$ and $\tilde{v}_{z}$ as contour maps in Figure 2.
We use the Kent and Bahcall models as representative models for the bulge and
disk
density distributions.
All the computations hereafter are carried out with density and velocity
distributions
toward Baade's Window unless another window is specified.
The isofrequency contours for a rotating bulge model are drawn with solid
lines,
while the dotted lines represent contours for a  nonrotating bulge model.
The rotation of the bulge does not dramatically affect the distribution of
$\Gamma$.
The event rates as functions of $\tilde{v}_{y}$ and $\tilde{R}_{e}$
are shown in Figure 3.
Note that very few events generated by disk stars have $\tilde{v}_{y} < 0$,
whereas those generated by bulge
stars are about equally likely to have negative as positive $\tilde{v}_{y}$.
Therefore, the reduced velocity $\tilde{ {\bf v}  }$ could be used for
partial separation of bulge from disk Machos.
Recall, however, that measurement of $\tilde{\bf {v}}$ generally requires
an additional satellite.

     Fortunately, the events generated by disk and bulge Machos can also
be distinguished by the reduced speed which generally requires only one
satellite.
In Figure 4, the rates of events with respect to $\tilde{v}$ and
$\tilde{R}_{e}$ are shown for bulge and disk Machos.
Note that the plotted values of $\tilde{R}_{e}$ are for an assumed
mass $M = 0.1\ {\rm M_{\odot}}$.
For other masses $\tilde{R}_{e} \propto M^{1/2}$.
The rates with respect to $\tilde{v}$ only are shown in Figure 5
for various models listed in Table 1.
For the spherical bulge model (and for either constant or linear disk velocity
dispersion model) the majority of the events lensed by the disk Machos would
occur in a region $\tilde{v} \leq 500\ {\rm km\ s^{-1}}$ while bulge Macho
lensing events would be found in a wide range of
$\tilde{v}\ (0-3000\ {\rm km\ s^{-1}})$.
This means that a significant fraction of the bulge Machos could be separated
from
the disk Machos: events with $\tilde{v} \leq 150\ {\rm km\ s^{-1}}$ are likely
to be
due to disk Machos whereas bulge Machos would be responsible for the events
with $\tilde{v} \geq 600\ {\rm km\ s^{-1}}$.
However, especially if the bulge has a bar-like structure
it would be hard to distinguish disk from bulge Machos
for $150 \lsim \tilde{v} \lsim 600\ {\rm km\ s^{-1}}$, i.e. a substantial
fraction of events [Fig.\ 5 (b)].
This predicament could be largely resolved by observing
stars through other windows toward galactic bulge.
The event rate as a function of $\tilde{v}$ through the window at
$l =  -8^{\circ}\hskip-2pt .66,\ b = -5^{\circ}\hskip-2pt .98$
(Blanco \& Terndrup 1989) is shown in Figure 5 (d).

\section{Mass Spectrum Of Machos}

   The reduced velocity $\tilde{v}$ determined from the parallax measurement of
lensing events could tell one the physical parameters of the Machos which
is very difficult to obtain from present observations.

     First, the approximate locations of disk Machos could be determined from
the reduced velocity measurements.
Assume for the moment that we could discriminate between disk and bulge Machos
from
the determined $\tilde{v}$.
Figure 6 shows the event rates with respect to $D_{ol}$ for different values of
$\tilde{v}$: the curves represent $\Gamma (D_{ol})$ for $\tilde{v} =
50,\ 100,\ 150$, and $300\ {\rm km\ s^{-1}}$ from the left to the right,
respectively.
Here, the event rates are computed with the linear disk velocity dispersion
model.
For each value of $\tilde{v}$, the most probable events occur at different
values of $D_{ol}$ enabling one to determine the approximate locations of
the disk Machos.
The assumption that $\tilde{v}$ discriminates between bulge and disk Machos
is valid over different ranges of $\tilde{v}$ for different windows.
For example, for Baade's Window separation can be made for
$\tilde{v} \lsim 150\ {\rm km\ s^{-1}}$, while for the
$l = -8^{\circ}\hskip-2pt.7$, $b = -6^{\circ}$ window, separation requires
$\tilde{v} \lsim 250\ {\rm km\ s^{-1}}$.
See Fig. 5.

     The next application of the reduced velocity measurement is
that it could provide a strong constraint on the mass spectrum of Machos.
Once again, let us assume that disk Machos could be distinguished from bulge
Machos.
The only constraint one has up to now for the estimate of Macho masses
is the charcteristic time $\omega^{-1}$.
For demonstration, say that a particular event has $\omega^{-1} = 10$ days,
which is the most common value measured by the MACHO group.
Figure 7 (a) shows the probability distribution of masses of a
disk Macho constrained by a single piece of information,
$\omega^{-1} = 10\ {\rm days}$,
while the probability distribution  constrained by the additional information,
$\tilde{v}$, is shown in Figure 7 (b).
The typical standard deviation of the distribution with two constraints
is $\sigma_{\log M} \sim 0.2$, whereas $\sigma_{\log M} \sim 0.6$ just
using the information of $\omega^{-1}$.
The additional information of $\tilde{v}$ significantly reduces the uncertainty
of
the mass estimate, and thus strongly constrains the mass spectrum of Machos.

     Unfortunately, it is more difficult to measure the mass spectrum
of bulge Machos.
The reduced Einstein ring radius and velocity can be approximated as
$\tilde{R}^{2}_{e} \sim (4GM/c^{2}) (R_{0}^{2}/D_{ls})$ and
$\tilde{v} \sim v (R_{0}/D_{ls})$.
Then,
$$
{M \over v} = {c^{2} \over 4GR_{0}} \left( {\tilde{v} \over
\tilde{R}_{e}^{2}} \right) \qquad ({\rm bulge}\ {\rm Machos}).
\eqno(4.2.1)
$$
The ratio $M/v$ can be determined from the measured value of $\tilde{R}_{e}$
and
$\tilde{v}$, but one needs to know about $v$ to obtain $M$.
This can be done only by careful modeling of the velocity distribution of
bulge Machos.

\section{Proper Motions  }

     For the lensing events in which a lens passes near the face of a star,
the light curve deviates from its characteristic 3 parameter curve [eq.\ (1.1)]
due to the differential magnification of the lensed star.
By using this principle, one could measure the proper motion
of a lens (Gould 1994a, Nemiroff \& Wickramasinghe 1994), which is defined by
$ \Omega = v / D_{ol} $.
In Figure 8, we plot the rate of lensing events as a function of $\Omega$.
The lensing events both by the disk and bulge Machos would occur
most probably around $ \Omega \sim 1.1\ {\rm \Omega_{0}}$,
where ${\rm \Omega_{0}} = v_{\odot}/R_{0}$.
Thus, it is unlikely that one could distinguish bulge Machos from disk Machos
by measuring of $\Omega$.
In addition, this measurement can be carried out for only $\sim 2
\ ({\rm M/0.1\ M_{\odot}})^{-1/2} \%$ of events.

     There would also be a detectable spectral line shift due to the rotation
of a lensed star.
By combining the line shift with $\omega^{-1}$ (from the overall light
curve) and with the stellar rotation speed (obtained from the spectrum),
the lens proper motion could be measured (Maoz \& Gould 1994).
Almost all the bulge stars observed by OGLE and MACHO have
color $V-I \gsim 0.83$ which are approximately K0 for main sequence stars
and G5 for giants.
Herbig $\&$ Spalding (1953, 1955) have shown observationally that there is a
cutoff in the distribution of rotation velocities for evolved stars,
which seem to be present near G0 in all luminosity classes
IV, III, II, and Ib.
Gray (1991) also provides a distribution of $v_{rot}$ .
In his distribution the drop in rotation of the giants takes place between
G0III and G3III.
Since nearly all lensing events occur later than the cutoff,
it would be difficult to detect the line shift due to the slow
rotation of lensed stars.
Moreover, even if the shift were detected, one must know the rotation speed
in order to measure the proper motion.
The rotation speed is difficult to measure when it is of order or less than the
turbulent velocities of the stars.

     It is important to note, however, that one can uniquely determine $v$,
$M$,
and $D_{ol}$ of Machos if both the proper motions and parallaxes are measured.
The distance to a Macho is
$$
D_{ol} = D_{os} \left[ \left({\Omega \over \tilde{v}} \right) D_{os} +1
\right]^{-1}.
\eqno(4.3.1)
$$
Since $D_{os} \sim 8\ {\rm kpc}$, one can determine $D_{ol}$ from measured
values of $\tilde{v}$ and $\Omega$.
Once $D_{ol}$ is known, the mass and transverse speed of a Macho can
be uniquely determined.

\chapter{Conclusions}

     The results of the computations of the optical depths and rates of lensing
events
with respect to various potentially measurable physical parameters are
summarized
as follows:

\noindent
1. The optical depths and rates of disk lensing events do not for the most part
depend strongly on the models that have been proposed, based on star counts
and stellar dynamics.
However, both the optical depths and event rates would be considerably  higher
if the bulge is barred.

\noindent
2. The events lensed by bulge Machos could be distinguished
from the ones lensed by disk Machos by measuring the reduced speed of Machos,
which can be done with a single satellite.

\noindent
3. The approximate locations and masses of a significant fraction of Machos
could be determined from the reduced speed of the Machos.
Therefore, $\tilde{v}$ would provide an important clue to the physical
distribution and mass spectrum of Machos.

\noindent
4. It would be difficult to measure the proper motions of Machos seen toward
the bulge either spectroscopically or photometrically.
Even when lens proper motions can be determined, they do not discriminate
well between bulge and disk Machos.
However, if both the proper motion and parallax of an event are measured, then
the mass, distance, and speed of the Macho are uniquely determined.

\endpage

\ref{Alcock, C. et al. 1994, ApJ, submitted}
\ref{Bahcall, J. N. 1986, ARA\&A, 24, 577}
\ref{Binney, J., \& Tremaine, S. 1987, Galactic Dynamics (princeton University
Press,
Princeton), 67}
\ref{Blanco, V. M., \& Terndrup, D. M. 1989, AJ, 98, 843}
\ref{Dwek, E., Arendt, R. G., Hauser, M. G., Kelsall, T., Lisse, C. M.,
Moseley, S. H., Silverberg, R. F., Sodroski, T. J., \& Weiland, J. 1994, ApJ,
submitted}
\ref{Evans, N. W. 1994, preprint}
\ref{Gould, A. 1994a, ApJL, 421, L71}
\ref{--------. 1994b, ApJL, 421, L75}
\ref{--------. 1994c, ApJL, 423, L105}
\ref{--------. 1994d, ApJL, submitted}
\ref{Gray, D. F. 1991, Angular Momentum Evolution of Young Stars
(Kluwer Academic Publishers, Netherland), 183}
\ref{Griest, K. et al. 1991, ApJ, 387, 181}
\ref{Han, C. 1994, in preparation}
\ref{Herbig, G. H., \& Spalding, J. F. 1953, PASP, 65, 192}
\ref{Herbig, G. H., \& Spalding, J. F. 1955, ApJ, 121, 118}
\ref{Kent, S. M., Dame, T. M., \& Fazio, G. 1991, ApJ, 378, 131}
\ref{Kent, S. M. 1992, ApJ, 387, 181}
\ref{Kiraga, M, \& Paczy\'nski, B. 1994, ApJL, in press}
\ref{Lewis, J. R., \& Freeman, K. C. 1989, AJ, 97, 139}
\ref{Maoz, D., \& Gould, A. 1994, ApJL, 425, L67}
\ref{Nemirnoff, R. J., \& Wickramasinghe, W. A. D. T. 1994, ApJ, 424, L21}
\ref{Paczy\'nski, B., Stanek, K., Z., Udalski, A.,
Szyma${\rm \acute{n}a}$ski, M., Kalu\.zny, J., Kubiak, M., \&
Krzemi${\rm \acute{n}a}$ski, W. 1994, ApJL, submitted}
\ref{Udalski, A., Szyma\'nski, J., Kalu\.zny, J., Kubiak, M., \& Krzemi\'nski,
W.
1994, ApJL, 426, L69}

\refout
\endpage
\endpage
\bye